\documentclass[journal=jacsat,manuscript=article]{achemso}
\setkeys{acs}{maxauthors=10, etalmode=truncate}

\usepackage{siunitx}
\usepackage{amsmath}     
\usepackage{amssymb}     
\usepackage{xcolor}
\usepackage[english]{babel} %Remark: use the same language for uvamath and babel
\usepackage{physics}
\usepackage{graphicx}
\usepackage[pdfborder={0 0 0}]{hyperref}
\usepackage{lipsum}
\usepackage{amsmath}
\usepackage{amsfonts} 
\usepackage{amsthm}
\usepackage{changepage}
\usepackage{graphicx, eso-pic, tikz}
\usepackage[utf8]{inputenc}
\usepackage[T1]{fontenc}
\usepackage{siunitx}
\usepackage{subcaption}
\usepackage{multirow}
\usepackage[version=4,arrows=pgf-filled,
textfontname=sffamily,
mathfontname=mathsf]{mhchem}
\usepackage{cleveref}
\author{Tijn Vernooij}
\affiliation{
    Advanced Research Center for Nanolithography (ARCNL), Science Park 106, 1098 XG Amsterdam, The Netherlands
}
\alsoaffiliation{
    Institute for Theoretical Physics, Institute of Physics, University of Amsterdam, Science Park 904, 1090 GL Amsterdam, The Netherlands
}

\author{H. Tun\c{c} \c{C}ift\c{c}i}
\affiliation{
    Advanced Research Center for Nanolithography (ARCNL), Science Park 106, 1098 XG Amsterdam, The Netherlands
}

\author{Noushine Shahidzadeh}
\affiliation{
    Van der Waals-Zeeman Institute, Institute of Physics, University of Amsterdam, Science Park 904, 1098 XH Amsterdam, The Netherlands
}

\author{Bart Weber}
\email{b.weber@arcnl.nl}
\affiliation{
    Advanced Research Center for Nanolithography (ARCNL), Science Park 106, 1098 XG Amsterdam, The Netherlands
}
\alsoaffiliation{
    Van der Waals-Zeeman Institute, Institute of Physics, University of Amsterdam, Science Park 904, 1098 XH Amsterdam, The Netherlands
}

\title{Amorphous \ce{CaSO4} nanocrystal deposits for friction and wear reduction at silicon interfaces}

\begin{document}

\maketitle

%%%%%%%%%%%%%%%%%%%%%%%%%%%%%%%%%%%%%%%%%%%%%%%%%%%%%%%%%%%%%%%%%%%%%
%% Abstract
%%%%%%%%%%%%%%%%%%%%%%%%%%%%%%%%%%%%%%%%%%%%%%%%%%%%%%%%%%%%%%%%%%%%%
\begin{abstract}

When an object is placed on a surface, friction and wear cause uncertainty in its exact position, and thus challenge precision manufacturing. Here, we explore the development of a sacrificial nanocrystal deposit that can suppress friction and wear. Amorphous \ce{CaSO4} nancrystals are deposited through salt solution droplet deposition followed by evaporation. During droplet drying, a precursor film of the aqueous \ce{CaSO4} solution spreads onto a hydrophilic silicon wafer, thus nucleating evenly spread unfaceted amorphous nanocrystals of \ce{CaSO4} on the wafer surface.  We used atomic force microscopy to study the extent, topography, and friction and wear behavior of the deposited nanocrystals. We find that the sacrificial layer of nanocrystals is easy to apply and remove, spreads over large (few cm) areas with a constant thickness of about 8 nm, and has favorable friction and wear behavior.

\end{abstract}

\newpage

%%%%%%%%%%%%%%%%%%%%%%%%%%%%%%%%%%%%%%%%%%%%%%%%%%%%%%%%%%%%%%%%%%%%%
%% Introduction
%%%%%%%%%%%%%%%%%%%%%%%%%%%%%%%%%%%%%%%%%%%%%%%%%%%%%%%%%%%%%%%%%%%%%

\section{Introduction}
In many areas of science and engineering, the ability to precisely locate samples on a positioner surface relative to a reference point is vital to maintaining experimental precision and repeatability \cite{precisionpositioning}$^,$\cite{precisionpositioning2}. When (sub)nanometer-scale precision is needed, friction and wear at the interface between the sample and the positioner can cause significant challenges\cite{burlpatent}. 
During repeated placement and removal of samples on a positioner surface, wear of the positioner may occur, even when the positioner surface is harder than the sample \cite{LERICHE2023204975}. Wear can cause uncertainty in the exact location and orientation of subsequent samples placed on the same positioner surface, reducing positioner performance and lifetime.
    
Hard ceramics are currently the standard choice for components that need to be highly resistant to wear \cite{oilgas}\cite{ceramictools}\cite{diamondtip}. However, even the hardest ceramics wear after repeated contact with softer materials\cite{hairsteel}\cite{LERICHE2023204975}\cite{underreviewbart}, such as silicon. Therefore, alternative or additional approaches towards wear suppression are desired.
    
Salt crystals can be of interest in the context of friction and wear control because salt crystals are stiff and can be easily grown and removed.
The crystalization pressure of \ce{NaCl} exceeds the tensile strength of many types of porous stones. As a result, salt crystallization is a major pathway for the degradation of rock and man-made structures\cite{artwork}\cite{egypt}\cite{AFMcrystallizationpressure}\cite{stonecrystallization}. At the same time, sliding on \ce{NaCl} and similar salts can involve relatively low friction coefficients of around 0.1 \cite{Saltfriction2}\cite{saltstickslip}\cite{earthquake}. Materials reinforced with \ce{CaSO4} whiskers generally have lower friction and higher wear resistance \cite{whiskersInPlastic}\cite{carbrakes}, while solid \ce{CaSO4} performs well as a high temperature lubricant \cite{aircraft}\cite{John1999}. At the same time, ceramic materials wear slower when sliding against lower hardness materials, such as salts \cite{YAMAMOTO199421}\cite{WANG1996112}. Gypsum has an indentation hardness of just $1\text{ to }\SI{1.5}{\giga Pa}$\cite{gypsum1}\cite{gypsum2} compared to about $\SI{10}{\giga Pa}$ for silicon\cite{LERICHE2023204975}. All these previous studies suggest that deposits made of relatively softer salts can protect a ceramic positioner surface from wear while also lowering friction.

Here, we investigate if a deposit of \ce{CaSO4} nanocrystals \cite{dropletevaporation} applied to naturally oxidized, polished silicon surfaces can reduce friction and wear at the interface between silicon surfaces. In order to grow \ce{CaSO4} nanocrystals on the silicon surface, a droplet  (V=$\SI{1}{\micro\liter}$) of nearly saturated \ce{CaSO4} solution is deposited at the oxygen plasma cleaned surface, followed by drying at $T=\SI{21}{\celsius}$ and RH=30-50\%.
The evaporation, which leads to crystallization of the salt, results in a homogeneous deposit of amorphous \ce{CaSO4} nanocrystals in the precursor wetting film of the doplet\cite{percussorfilm}\cite{creepingtube}. The nanocrystals formed in this confined thin film are unfaceted with a flattened spherical shape.
We demonstrate that silicon atomic force microscope tips sliding over \ce{CaSO4} crystals experience 40\% lower friction and 70\% lower wear compared to sliding directly on silicon. 

%%%%%%%%%%%%%%%%%%%%%%%%%%%%%%%%%%%%%%%%%%%%%%%%%%%%%%%%%%%%%%%%%%%%%
%% Methods
%%%%%%%%%%%%%%%%%%%%%%%%%%%%%%%%%%%%%%%%%%%%%%%%%%%%%%%%%%%%%%%%%%%%%

\section{Methods}

\subsection{Creation and removal of the deposit}

As a substrate, we used polished, naturally oxidized, p-doped silicon wafers (University Wafer; $<$100$>$ orientation; boron-doped; single-side polished; 500–525~$\mu$m thick; 1–10~$\Omega\cdot$cm resistivity). We cut the wafers into pieces of approximately $2 \times 2\ \SI{}{\centi\meter\squared}$. We cleaned the substrates by sequentially rinsing them with deionized water, hot tap water, ethanol, and deionized water again. We then plasma cleaned the samples for 10 minutes using oxygen gas in a Diener Electronic Zepto plasma cleaner. We placed a single $\SI{1}{\micro\liter}$ water droplet with a \ce{CaSO4} concentration of $\SI{1.9}{g/L}$ on each plasma-cleaned sample. Upon evaporation—typically within about one minute at a relative humidity of 30–50\%—the droplet left behind the \ce{CaSO4} crystal deposit of interest on the surface. 

\subsection{Topography measurement}

We performed topography measurements of the \ce{CaSO4} deposit using both scanning electron microscopy (SEM; FEI Verios 460) and tapping mode atomic force microscopy (AFM; Bruker Dimension Icon and Bruker Innova).  

\subsection{Friction coefficient measurement}

To compare the friction coefficients of the \ce{CaSO4} crystals and uncoated silicon surfaces, we used lateral force microscopy (LFM)~\cite{lfm1}. We acquired lateral deflection images that included both \ce{CaSO4} crystals and the surrounding silicon wafer surface not covered by crystals. We subtracted the forward and backward scan directions to create friction coefficient maps in arbitrary units~\cite{lfm2}. We applied this method to all three regions (the droplet edge, the transition region, and the precursor film of nanocrystals) of the deposited \ce{CaSO4}. In this way, we compared the friction coefficient of \ce{CaSO4} crystals—ranging in size from $\SI{20}{\nano\meter}$ to $\SI{10}{\micro\meter}$—to the smooth surface of naturally oxidized silicon. For these measurements, we used Bruker RESPA-20 and RESPA-10 model tips.

\subsection{Tip wear measurement}

In order to evaluate whether the nanocrystal deposit could protect a ceramic positioner surface from wear, we measured the wear rate of Bruker RTESPA 150 model silicon AFM tips when sliding against \ce{CaSO4} crystals compared to sliding against uncoated p-type silicon wafers. We followed a protocol suited for wear tests on flat surfaces~\cite{quantifytipwear}; First, we performed repeated contact mode scans with a normal force of $\SI{100}{\nano\newton}$ over a $1 \times 1~\SI{}{\micro\meter\squared}$ area, which led to wear of a pyramidal section of the tip. We then imaged a Bruker RS-12M roughness sample using these degraded tips. The triangular base of the worn tip appeared as repeating features on the sample peaks. Based on the known tip geometry, we determined that the worn tip volume is equal to $0.21 \cdot (\text{height of triangular base})^3$. This derivation is provided in the Supplementary Information. Using this approach, we calculated the volume of the worn pyramid-shaped tip section and, from that, computed the average tip wear rate in $\SI{}{\micro\meter^3/(N\cdot m)}$ for both uncoated silicon and the large gypsum crystals present at the \textit{droplet edge}. Gypsum crystals were used in the wear experiments because these are the only \ce{CaSO4} crystals large enough to accommodate the needed $1 \times 1~\SI{}{\micro\meter\squared}$ contact mode scan area.

\subsection{Resilience of individual nanocrystals}

Displacement of nanocrystals at frictional contacts may either accommodate relative motion (desirable) or cause contamination through nanocrystal redistribution (undesirable). To investigate the normal forces required to displace individual nanocrystals, we performed $3 \times 3~\SI{}{\micro\meter^2}$ contact mode AFM scans with normal forces ranging from 30 to 300~$\SI{}{\nano\newton}$, followed by tapping mode topography imaging to assess the resulting nanocrystal displacement.

%%%%%%%%%%%%%%%%%%%%%%%%%%%%%%%%%%%%%%%%%%%%%%%%%%%%%%%%%%%%%%%%%%%%%
%% New Results
%%%%%%%%%%%%%%%%%%%%%%%%%%%%%%%%%%%%%%%%%%%%%%%%%%%%%%%%%%%%%%%%%%%%%

\section{Results}

\subsection{Deposition morphology and zonal structure}

\begin{figure}[h!]
     \centering
     \includegraphics[width=\linewidth]{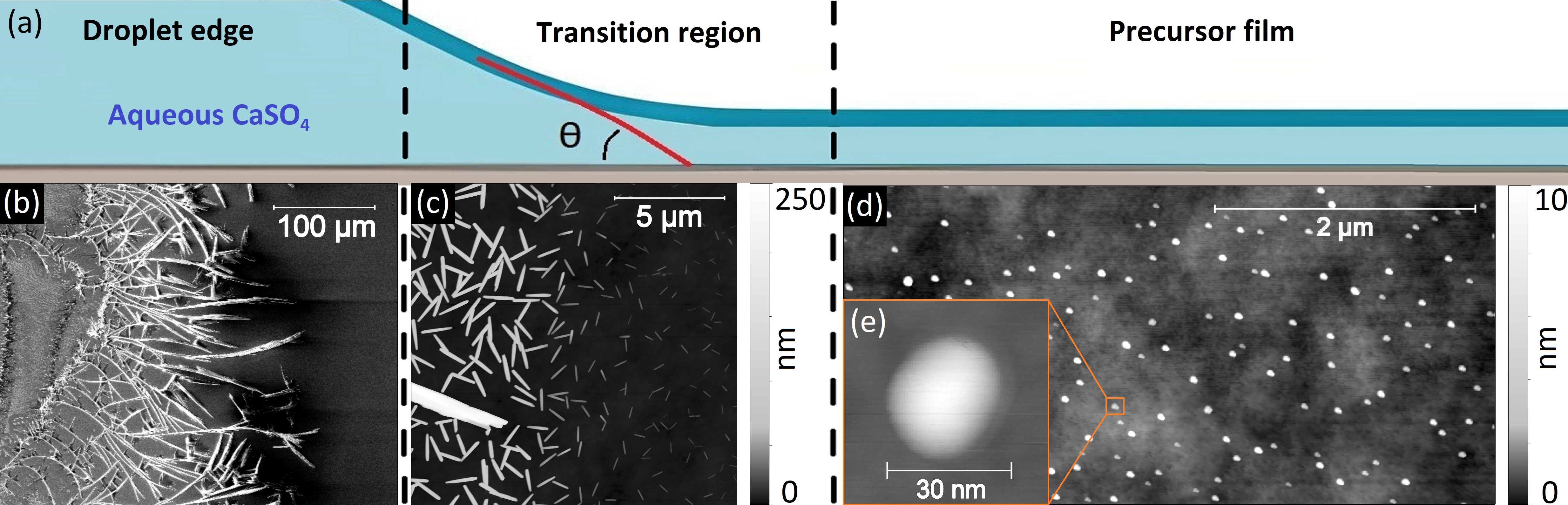}
     \caption{(a) Schematic of a droplet of aqueous \ce{CaSO4} solution evaporating on a silicon substrate, with a macroscopic contact angle $\SI{0}{\degree} < \theta < \SI{2}{\degree}$ and a precursor film extending several centimeters from the droplet edge. Upon evaporation, three distinct deposition zones form, each characterized by different \ce{CaSO4} morphologies. (b) SEM image of gypsum crystals (\ce{CaSO4}$\cdot$2\ce{H2O}) marking the former macroscopic contact line at the droplet edge. AFM images of (c) bassanite crystals (\ce{CaSO4}$\cdot\frac{1}{2}$\ce{H2O}) in the intermediate \textit{transition region} between the droplet edge and the precursor film, (d) an amorphous nanocrystal deposit (\ce{CaSO4}$\cdot x$\ce{H2O}, water content $x>0$ varies\cite{caso4crystalization1,caso4crystalization2,caso4crystalization3,caso4crystalization4,ACSwater}) within the precursor film region, and (e) a close-up of an individual nanocrystal. Additional images are provided in the Supplementary Information.}
     \label{topography}
\end{figure}

We achieved dense and spatially extended nanocrystal deposits through droplet evaporation of aqueous \ce{CaSO4}, as shown in SEM and AFM scans in Figure~\ref{topography}. The rapid evaporation at the droplet's edge generates a localized ion concentration gradient, prompting crystallization near the contact line. Nevertheless, crystal morphology varies significantly with location, allowing us to identify three distinct zones: a circular \textit{droplet edge} that traces the initial three-phase contact line of the evaporated droplet (Figure~\ref{topography}b); an intermediate \textit{transition region} where the crystal size gradually decreases (Figure~\ref{topography}c); and a homogeneous \textit{nanocrystal deposit} extending outward from the edge (Figure~\ref{topography}d). As our study focuses primarily on the nanocrystal region, we provide only a brief overview of the droplet edge and transition region here, with further details available in the Supplementary Information.

The structure of the droplet edge varied between and within samples. Some edges consisted of widely spaced gypsum crystals (Figure~\ref{topography}b), while others featured a compact layer of fine bassanite crystals that completely covered the substrate (see Supplementary Information). At bassanite-rich edges, we observed a gradual reduction in nanocrystal size leading into the transition region (Figure~\ref{topography}c), eventually merging with the amorphous nanocrystal zone (Figure~\ref{topography}d). In contrast, gypsum-dominated edges exhibited abrupt transitions to the nanocrystal zone, with no intermediate-sized crystals (Figure~S2). These differences arise because the low thickness of the liquid film near the droplet edge imposes geometric constraints on the crystalization of \ce{CaSO4}, with different polymorphs being favored depending on the size of the crystal that can form.

The homogeneous nanocrystal region extended at least \SI{5}{\centi\meter} from the droplet edge with minimal variation in morphology or density (Figures~\ref{topography}d and S5). AFM scans revealed that individual nanocrystals formed flattened spherical caps with heights of approximately \SI{8}{\nano\meter} and lateral dimensions of \SI{20}{\nano\meter} (Figure~\ref{topography}e). Based on surface coverage analysis, we estimated that 1–5\% of the wafer area was directly covered by nanocrystals. Considering AFM tip convolution, we used 1\% as a conservative estimate, corresponding to a number density of approximately \SI{6e13}{\per\square\meter}, or an average center-to-center spacing of roughly \SI{125}{\nano\meter}.

\subsection{Mechanical stability and removability of the nanocrystal layer}

To measure how firmly the nanocrystals adhere to the wafer surface, we performed contact-mode AFM scans with normal forces ranging from 30 to \SI{300}{\nano\newton}. We found that individual nanocrystals remained in place under forces up to approximately \SI{30}{\nano\newton}, but were displaced at higher loads. Phase contrast images (Figure~S6b) confirmed that displaced nanocrystals left no detectable residue at their original locations. This could be explained by the unfaceted shape of the nanocrystals, allowing them to roll over the surface when force is applied.

Given the nanocrystal number density of \SI{6e13}{\per\square\meter}, we estimated that the deposit can withstand macroscopic pressures of up to \SI{2}{\mega\pascal} before widespread nanocrystal displacement occurs (Figure~S6). This threshold is substantially higher than the typical clamping pressures of \SI{0.1}{\mega\pascal}\cite{chuck} used in vacuum chuck systems in the semiconductor industry, suggesting that the deposit may remain stable under practical conditions.

In addition, to test how easily the layer can be removed for cleaning purposes, we rinsed and immersed samples in distilled water. In all cases, the \ce{CaSO4} layer dissolved completely within one minute, leaving no observable residue on the silicon surface (Figure~S8). These results demonstrate that the nanocrystal deposit is both easy to create and easy to remove, which makes it attractive as a sacrificial spacer at (self-mated silicon) tribological interfaces.

\subsection{Friction coefficients of \ce{CaSO4} crystals and nanocrystals}

\begin{figure}[h!]
     \centering
     \begin{subfigure}[b]{0.32\textwidth}
         \centering
         \includegraphics[width=\textwidth]{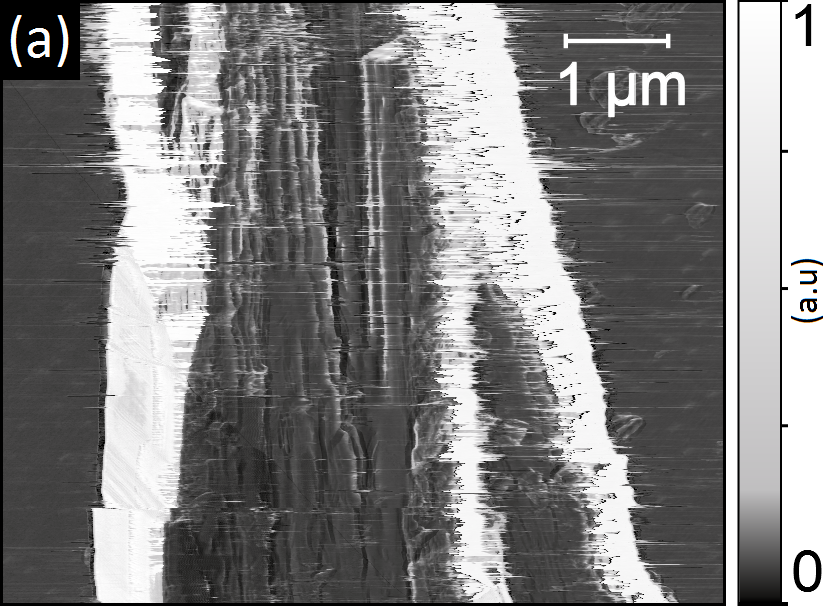}
%         \caption{$y=x$}
         \label{fig:y equals x}
     \end{subfigure}
     \begin{subfigure}[b]{0.32\textwidth}
         \centering
         \includegraphics[width=\textwidth]{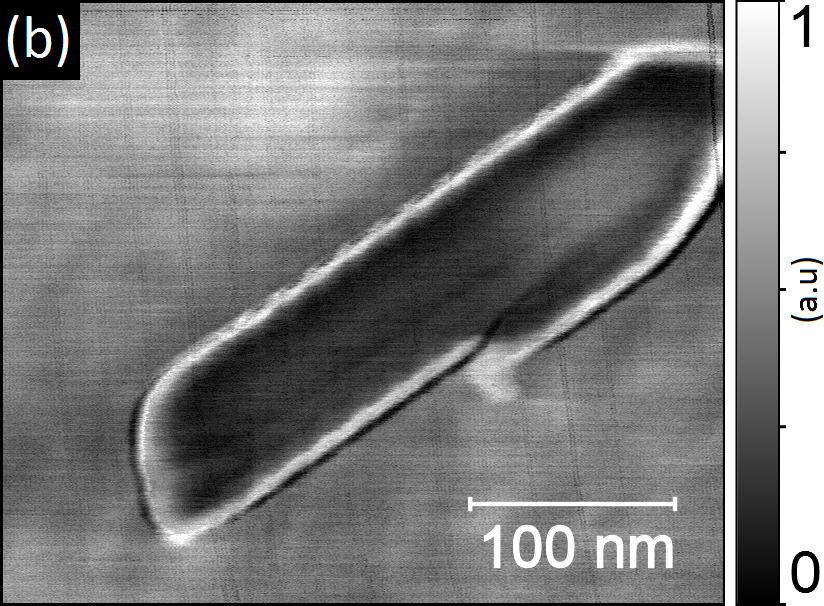}
%         \caption{$y=3\sin x$}
         \label{fig:three sin x}
     \end{subfigure}
     \begin{subfigure}[b]{0.32\textwidth}
         \centering
         \includegraphics[width=\textwidth]{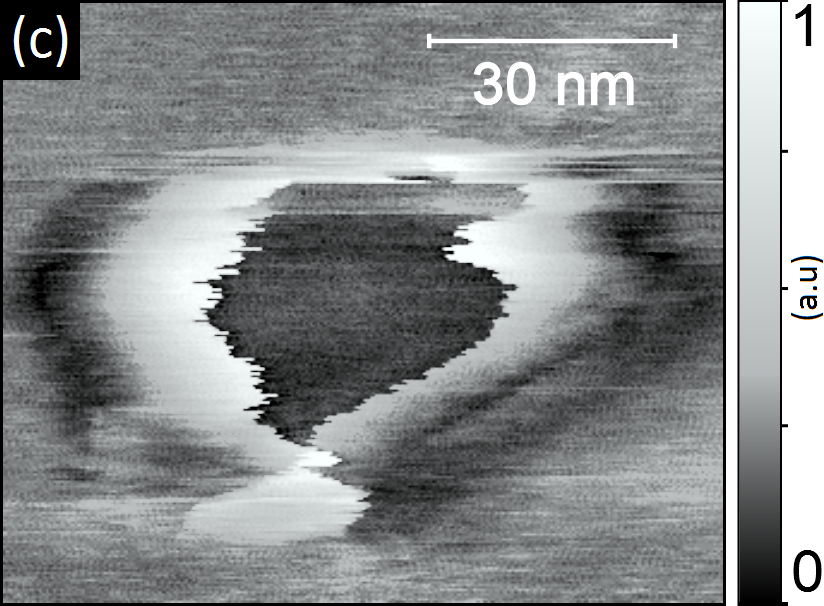}
%         \caption{$y=3\sin x$}
         \label{fig:three sin z}
     \end{subfigure}
        \caption{Friction coefficient maps (in arbitrary units, a.u.) of three types of \ce{CaSO4} deposits: (a) a gypsum crystal from the droplet edge, (b) a bassanite crystal from the transition region, and (c) an amorphous nanocrystal from the precursor film area.}
        \label{friction}
\end{figure}

In Figure~\ref{friction}, lateral force microscopy (LFM) maps of three types of \ce{CaSO4} morphologies are presented: (i) large gypsum crystals at the droplet edge (Figure~\ref{friction}a), (ii) bassanite crystals in the transition region (Figure~\ref{friction}b), and (iii) amorphous nanocrystals from the precursor film region (Figure~\ref{friction}c). 

Each LFM pixel intensity reflects the lateral force measured at that point. Boundaries of individual crystals appear as high-friction zones because the tip experiences resistance as it climbs on top the crystal. By comparing the average friction on top of the crystals with the
average values outside the crystals, we estimate the ratio of the friction coefficient of silicon
and the three types of CaSO4 crystals. In this way, we find that the friction coefficient of \ce{CaSO4} nanoparticles and larger bassanite crystals is approximately 40\% lower than that of
the surrounding uncoated silicon. This observation suggests that the nanoparticle deposit
effectively reduces the friction coefficient of the silicon surface to which it is applied. The
friction coefficient of the large gypsum crystal is strongly variable. On flat regions the friction coefficient is about 40\% lower than the surrounding silicon, just like for nanoparticles and bassanite crystals, but in many other places it is much higher. This can be attributed to the high surface roughness of large gypsum crystals in comparison to both uncoated silicon and the other types of \ce{CaSO4} crystals.

\subsection{AFM tip wear on silicon and gypsum crystals}

\begin{figure}[h!]
    \centering
    \includegraphics[width=\linewidth]{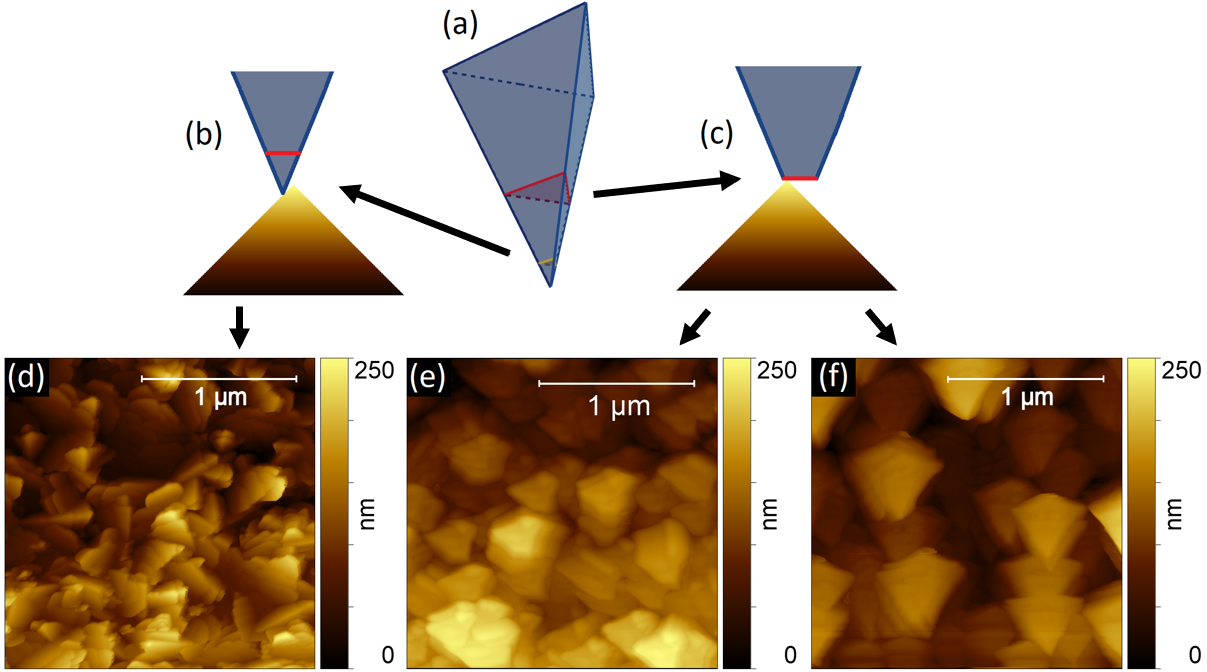}
    \caption{Procedure used to estimate the worn volume of an AFM tip. (a) Model of the RTESPA-150 tip used in experiments. As the tip slides over a flat surface, material is gradually removed from the apex, exposing a triangular cross-section that increases in size with wear—starting small (yellow surface) and growing larger (red surface). (b,c) Schematics of a lightly worn tip (b) and a heavily worn tip (c) scanning a sharp feature of the reference roughness sample. Because the sample features are sharper than the worn tips, the resulting images effectively demonstrate the geometry of the degraded tip. (d–f) Scans of the Bruker RS-12M roughness sample acquired using (d) a fresh tip, (e) a tip worn on gypsum, and (f) a tip worn on silicon. The size of the exposed triangular cross-section is extracted from these images, and the corresponding wear volume is calculated. Additional details are provided in the Supplementary Information (Figures S9 and S10).}
    \label{tipwear}
\end{figure}

To quantify the degree to which the \ce{CaSO4} deposits can suppress silicon-on-silicon
wear, we performed controlled wear experiments (see Methods, Figure~\ref{tipwear} and Table 1) both on gypsum crystals located at the droplet edge (Figure~\ref{topography}b) and on uncoated silicon wafers. 

\begin{table}[h!]
\centering
\begin{tabular}{ |p{7cm}||p{3cm}|p{3cm}|  }
 \hline
 \multicolumn{3}{|c|}{Results of tip wear rate experiments on silicon and gypsum crystals} \\
 \hline
  & Si wafer & Gypsum crystal \\
 \hline
 Contact force ($\SI{} {\nano N}$)    &  100    & 100 \\
 Number of measurements &   5  & 10 \\
 Scan distances (mm) & 5--102 & 3--240 \\
 Average wear rate ($ \SI{e5 }{\micro m^3/(Nm)}$) &   5.2  & 1.5 \\
 Minimum and maximum wear rate ($ \SI{e5 }{\micro m^3/(Nm)}$) & 3.8--9.3 & 0.08--4.2 \\
 Uncertainty per measurement (\%) & 30  & 70  \\
 Uncertainty of average result (\%) & 14 & 22 \\
 \hline
\end{tabular}
\caption{Summary of the wear rate of pristine silicon AFM tips sliding over either pristine silicon wafers or gypsum crystals. The wear rate on softer~\cite{gypsum1,gypsum2} gypsum is substantially lower than on harder\cite{LERICHE2023204975} silicon.}
\label{table}
\end{table}

The measured silicon-on-silicon wear rate agrees with previous measurements of silicon wear under self-mated and ambient conditions\cite{Bartfriction}, thereby providing independent support for our methodology. On average, the tip wear rate was 70\% lower when scanning over gypsum (\SI{1.5e5}{\micro\meter^3/(N\meter)}) compared to silicon (\SI{5.2e5}{\micro\meter^3/(N\meter)}). This
suggests that the nanoparticle deposit can indeed significantly reduce wear of surfaces sliding
against a silicon sample, as may be expected based on the lower hardness of gypsum~\cite{gypsum1,gypsum2} compared to silicon. Individual measurements on gypsum exhibited higher uncertainty due to irregular wear profiles on rough crystal surfaces, whereas scans on smooth silicon showed less variation in average wear rate.

%%%%%%%%%%%%%%%%%%%%%%%%%%%%%%%%%%%%%%%%%%%%%%%%%%%%%%%%%%%%%%%%%%%%%
%% Discussion
%%%%%%%%%%%%%%%%%%%%%%%%%%%%%%%%%%%%%%%%%%%%%%%%%%%%%%%%%%%%%%%%%%%%%

\section{Discussion}

\subsection*{Nanocrystal distribution and interfacial applications}

The investigated nanocrystal deposit remains morphologically consistent over several centimeters, confirming the uniformity and scalability of the deposition approach.
This uniformity makes the deposit a promising candidate for use as a sacrificial interfacial layer in applications that require nanoscale separation between flat surfaces—such as between silicon wafer backsides and positioners in semiconductor equipment. The measured height and lateral dimensions of the nanocrystals are sufficient to maintain a controlled separation, potentially reducing direct contact and the resulting friction and wear in precision positioning systems.

Based on morphological comparison with literature~\cite{caso4crystalization1,caso4crystalization2}, we attribute the deposited nanocrystals to quasi spherical amorphous calclium sulfate, and the nanorod shapes to bassanite (\ce{CaSO4}$\cdot\frac{1}{2}$\ce{H2O}). The amorphous character may offer mechanical compliance and low adhesion, facilitating easy removal and minimizing contamination. These features could support the development of reversible surface separation strategies in cleanroom environments.

\subsection*{Stability and reversibility of the nanocrystal layer}

We found that individual nanocrystals withstand normal forces of up to approximately \SI{30}{\nano\newton}, corresponding to a macroscopic pressure of about \SI{2}{\mega\pascal}. This value exceeds the typical clamping pressures of \SI{0.1}{\mega\pascal}\cite{chuck} used in vacuum chuck systems, indicating that the deposit remains stable under standard operating conditions in semiconductor processing.

We also confirmed that the layer can be fully removed by simple rinsing or brief immersion in distilled water, without leaving detectable residues. This ease of removal highlights the reversibility of the deposition and supports its suitability as a temporary or disposable interfacial layer.

\subsection*{Friction reduction and morphology effects}

Our lateral force microscopy measurements showed that both amorphous \ce{CaSO4} nanocrystals and bassanite nanorod crystals reduce local friction coefficients by approximately 40\% compared to the surface of the bare silicon. This reduction is especially valuable in precision positioning systems, where friction-induced microdeformation and wear can degrade performance~\cite{frictionprecision}.

In contrast, larger gypsum crystals exhibited spatially heterogeneous friction behavior due to their faceted shape. In flat regions, friction was relatively low, but in rougher areas, friction values were significantly higher. We attribute this variation to increased local asperities and topographical irregularities. These results suggest that smoother morphologies, such as the quasi-spherical amorphous nanocrytals and nanorods of bassanite, offer more consistent and predictable frictional behavior, making them better suited for integration into precision tribological systems.

\subsection*{Wear reduction on gypsum surfaces}

Our AFM tip wear experiments demonstrated that sliding on gypsum crystals reduced tip wear rates by approximately 70\% compared to sliding on silicon. Despite the relatively rough surface of gypsum, this protective effect was consistently observed across multiple measurements.

This finding is significant for applications where uncoated silicon components undergo repeated sliding, such as in micro-positioning stages or MEMS devices. Reducing tip or surface wear with \ce{CaSO4} deposits could extend component lifetimes and improve reliability.

While tip wear measurements on gypsum showed higher uncertainty due to irregular wear geometries, the overall trend remains robust. The smoother surface of silicon allowed for more precise volume loss measurements and served as a reliable baseline for comparison. These results support the concept that sacrificial or compliant surface layers can mitigate wear in silicon-on-silicon contact scenarios.

\subsection*{Outlook}

We have shown that \ce{CaSO4} nanocrystals offer favorable friction and wear properties at the single-nanocrystal level. However, practical implementations will involve larger contact areas where many nanocrystals simultaneously contribute to interfacial behavior. A natural next step is to conduct macroscale friction experiments—such as wafer-on-wafer contact tests—to determine whether these beneficial effects persist at larger scales.

We selected \ce{CaSO4} for this study due to its previously reported ability to form stable nanocrystals during evaporation~\cite{dropletevaporation,caso4crystalization1,caso4crystalization2}. However, we see no fundamental reason to assume that \ce{CaSO4} performs better than other low solubility salts. Since the deposition method is general, future work could explore analogous deposits formed from other salt systems, particularly those already reported in literature~\cite{creepingtube}. Furthermore, the influence of other deposition methods or drying conditions could be studied.

Finally, we suggest calibrating the lateral force microscopy measurements using standard protocols~\cite{calibration1,calibration2} to convert friction maps from arbitrary units to absolute values. This calibration would allow direct quantitative comparisons with commercial friction-reducing coatings and help benchmark nanocrystal-based strategies in industrial contexts.

%%%%%%%%%%%%%%%%%%%%%%%%%%%%%%%%%%%%%%%%%%%%%%%%%%%%%%%%%%%%%%%%%%%%%
%% Conclusion
%%%%%%%%%%%%%%%%%%%%%%%%%%%%%%%%%%%%%%%%%%%%%%%%%%%%%%%%%%%%%%%%%%%%%

\section{Conclusion}

We demonstrated that droplet-based nanocrystal deposition offers a simple and effective method to form dense and spatially uniform nanocrystal layers on silicon wafers. These layers span several centimeters with consistent morphology and coverage, enabling their potential use as sacrificial interfacial coatings in flat-on-flat contact applications.

We characterized the deposited nanocrystals as quasi spherical amorphous calcium sulfate with average dimensions of approximately \SI{8}{\nano\meter} in height and \SI{20}{\nano\meter} in lateral diameter. The nanocrystals remained mechanically stable under normal forces up to \SI{30}{\nano\newton}, corresponding to a macroscopic pressure of about \SI{2}{\mega\pascal}, well above the pressures used in vacuum chuck systems. Additionally, the entire deposit could be removed easily by rinsing with water, leaving no observable residue.

Through lateral force microscopy, we found that both amorphous nanocrystals and bassanite crystals reduced surface friction by approximately 40\% compared to bare silicon. Wear tests using silicon AFM tips revealed a 70\% reduction in tip wear when sliding over gypsum crystals relative to silicon, further underscoring the deposit’s protective potential.

Together, these findings indicate that \ce{CaSO4} nanocrystal layers can reduce friction and wear at critical ceramic interfaces, such as those found in semiconductor positioning systems. Their ease of application, reversibility, and tribological performance make them promising candidates for integration into cleanroom-compatible, precision-engineered surfaces.

%%%%%%%%%%%%%%%%%%%%%%%%%%%%%%%%%%%%%%%%%%%%%%%%%%%%%%%%%%%%%%%%%%%%%
%% Acknowledgements
%%%%%%%%%%%%%%%%%%%%%%%%%%%%%%%%%%%%%%%%%%%%%%%%%%%%%%%%%%%%%%%%%%%%%
\begin{acknowledgement}

This work was conducted at the Advanced Research Center for Nanolithography, a public-private partnership between the University of Amsterdam (UvA), Vrije Universiteit Amsterdam (VU), Rijksuniversiteit Groningen (RUG), the Netherlands Organization for Scientific Research (NWO), and the semiconductor equipment manufacturer ASML.
We thank Ozan Sahin for making high-resolution SEM images of one of our samples. These images are used in Figure \ref{topography} and Figure S2 of the Supplementary Information.
%This work is based on a Master graduation project carried out at the Advanced Research Center for Nanolithography (ARCNL)

\end{acknowledgement}

%%%%%%%%%%%%%%%%%%%%%%%%%%%%%%%%%%%%%%%%%%%%%%%%%%%%%%%%%%%%%%%%%%%%%
%% References
%%%%%%%%%%%%%%%%%%%%%%%%%%%%%%%%%%%%%%%%%%%%%%%%%%%%%%%%%%%%%%%%%%%%%
\newpage

\bibliography{refs.bib}

\providecommand{\latin}[1]{#1}
\makeatletter
\providecommand{\doi}
  {\begingroup\let\do\@makeother\dospecials
  \catcode`\{=1 \catcode`\}=2 \doi@aux}
\providecommand{\doi@aux}[1]{\endgroup\texttt{#1}}
\makeatother
\providecommand*\mcitethebibliography{\thebibliography}
\csname @ifundefined\endcsname{endmcitethebibliography}  {\let\endmcitethebibliography\endthebibliography}{}
\begin{mcitethebibliography}{41}
\providecommand*\natexlab[1]{#1}
\providecommand*\mciteSetBstSublistMode[1]{}
\providecommand*\mciteSetBstMaxWidthForm[2]{}
\providecommand*\mciteBstWouldAddEndPuncttrue
  {\def\EndOfBibitem{\unskip.}}
\providecommand*\mciteBstWouldAddEndPunctfalse
  {\let\EndOfBibitem\relax}
\providecommand*\mciteSetBstMidEndSepPunct[3]{}
\providecommand*\mciteSetBstSublistLabelBeginEnd[3]{}
\providecommand*\EndOfBibitem{}
\mciteSetBstSublistMode{f}
\mciteSetBstMaxWidthForm{subitem}{(\alph{mcitesubitemcount})}
\mciteSetBstSublistLabelBeginEnd
  {\mcitemaxwidthsubitemform\space}
  {\relax}
  {\relax}

\bibitem[Gao \latin{et~al.}(2015)Gao, Kim, Bosse, Haitjema, Chen, Lu, Knapp, Weckenmann, Estler, and Kunzmann]{precisionpositioning}
Gao,~W.; Kim,~S.; Bosse,~H.; Haitjema,~H.; Chen,~Y.; Lu,~X.; Knapp,~W.; Weckenmann,~A.; Estler,~W.; Kunzmann,~H. Measurement technologies for precision positioning. \emph{CIRP Annals} \textbf{2015}, \emph{64}, 773--796\relax
\mciteBstWouldAddEndPuncttrue
\mciteSetBstMidEndSepPunct{\mcitedefaultmidpunct}
{\mcitedefaultendpunct}{\mcitedefaultseppunct}\relax
\EndOfBibitem
\bibitem[Oiwa \latin{et~al.}(2011)Oiwa, Katsuki, Karita, Gao, Makinouchi, Sato, and Oohashi]{precisionpositioning2}
Oiwa,~T.; Katsuki,~M.; Karita,~M.; Gao,~W.; Makinouchi,~S.; Sato,~K.; Oohashi,~Y. Questionnaire survey on ultra-precision positioning. \emph{International Journal of Automation Technology} \textbf{2011}, \emph{5}, 766--772\relax
\mciteBstWouldAddEndPuncttrue
\mciteSetBstMidEndSepPunct{\mcitedefaultmidpunct}
{\mcitedefaultendpunct}{\mcitedefaultseppunct}\relax
\EndOfBibitem
\bibitem[bur(2017)]{burlpatent}
Substrate holder and a method of manufacturing a substrate holder. 2017; US Patent US10719019B2\relax
\mciteBstWouldAddEndPuncttrue
\mciteSetBstMidEndSepPunct{\mcitedefaultmidpunct}
{\mcitedefaultendpunct}{\mcitedefaultseppunct}\relax
\EndOfBibitem
\bibitem[Leriche \latin{et~al.}(2023)Leriche, Xiao, Franklin, and Weber]{LERICHE2023204975}
Leriche,~C.; Xiao,~C.; Franklin,~S.; Weber,~B. From atomic attrition to mild wear at multi-asperity interfaces: The wear of hard Si3N4 repeatedly contacted against soft Si. \emph{Wear} \textbf{2023}, \emph{528-529}, 204975\relax
\mciteBstWouldAddEndPuncttrue
\mciteSetBstMidEndSepPunct{\mcitedefaultmidpunct}
{\mcitedefaultendpunct}{\mcitedefaultseppunct}\relax
\EndOfBibitem
\bibitem[Medvedovski(2018)]{oilgas}
Medvedovski,~E. \emph{Engineering ceramics for wear-protection of mining and mineral processing equipment}; 2018; pp 647--651\relax
\mciteBstWouldAddEndPuncttrue
\mciteSetBstMidEndSepPunct{\mcitedefaultmidpunct}
{\mcitedefaultendpunct}{\mcitedefaultseppunct}\relax
\EndOfBibitem
\bibitem[Li and Low(1994)Li, and Low]{ceramictools}
Li,~X.~S.; Low,~I.~M. Ceramic cutting tools-an introduction. \emph{Key Engineering Materials} \textbf{1994}, \emph{96}, 1--18\relax
\mciteBstWouldAddEndPuncttrue
\mciteSetBstMidEndSepPunct{\mcitedefaultmidpunct}
{\mcitedefaultendpunct}{\mcitedefaultseppunct}\relax
\EndOfBibitem
\bibitem[Bhaskaran \latin{et~al.}(2010)Bhaskaran, Gotsmann, Sebastian, Drechsler, Lantz, Despont, Jaroenapibal, Carpick, Chen, and Sridharan]{diamondtip}
Bhaskaran,~H.; Gotsmann,~B.; Sebastian,~A.; Drechsler,~U.; Lantz,~M.; Despont,~M.; Jaroenapibal,~P.; Carpick,~R.; Chen,~Y.; Sridharan,~K. Ultralow nanoscale wear through atom-by-atom attrition in silicon-containing diamond-like carbon. \emph{Nature nanotechnology} \textbf{2010}, \emph{5}, 181--5\relax
\mciteBstWouldAddEndPuncttrue
\mciteSetBstMidEndSepPunct{\mcitedefaultmidpunct}
{\mcitedefaultendpunct}{\mcitedefaultseppunct}\relax
\EndOfBibitem
\bibitem[Roscioli \latin{et~al.}(2020)Roscioli, Taheri-Mousavi, and Tasan]{hairsteel}
Roscioli,~G.; Taheri-Mousavi,~S.~M.; Tasan,~C.~C. How hair deforms steel. \emph{Science} \textbf{2020}, \emph{369}, 689--694\relax
\mciteBstWouldAddEndPuncttrue
\mciteSetBstMidEndSepPunct{\mcitedefaultmidpunct}
{\mcitedefaultendpunct}{\mcitedefaultseppunct}\relax
\EndOfBibitem
\bibitem[Leriche \latin{et~al.}(2025)Leriche, Pedretti, Kang, Righi, and Weber]{underreviewbart}
Leriche,~C.; Pedretti,~E.; Kang,~D.; Righi,~M.~C.; Weber,~B. Passivation species suppress atom-by-atom wear of micro-crystalline diamond. \emph{Under review} \textbf{2025}, \relax
\mciteBstWouldAddEndPunctfalse
\mciteSetBstMidEndSepPunct{\mcitedefaultmidpunct}
{}{\mcitedefaultseppunct}\relax
\EndOfBibitem
\bibitem[Desarnaud \latin{et~al.}(2016)Desarnaud, Bonn, and Noushine]{artwork}
Desarnaud,~J.; Bonn,~D.; Noushine,~S. Measurement of the pressure induced by salt crystallization in confinement. \emph{Scientific Reports} \textbf{2016}, \emph{6}\relax
\mciteBstWouldAddEndPuncttrue
\mciteSetBstMidEndSepPunct{\mcitedefaultmidpunct}
{\mcitedefaultendpunct}{\mcitedefaultseppunct}\relax
\EndOfBibitem
\bibitem[Wüst and Schlüchter(2000)Wüst, and Schlüchter]{egypt}
Wüst,~R.~A.; Schlüchter,~C. The origin of soluble salts in rocks of the Thebes Mountains, Egypt: The damage potential to ancient Egyptian wall art. \emph{Journal of Archaeological Science} \textbf{2000}, \emph{27}, 1161--1172\relax
\mciteBstWouldAddEndPuncttrue
\mciteSetBstMidEndSepPunct{\mcitedefaultmidpunct}
{\mcitedefaultendpunct}{\mcitedefaultseppunct}\relax
\EndOfBibitem
\bibitem[Hamilton \latin{et~al.}(2010)Hamilton, Koutsos, and Hall]{AFMcrystallizationpressure}
Hamilton,~A.; Koutsos,~V.; Hall,~C. Direct measurement of salt-mineral repulsion using atomic force microscopy. \emph{Chemical communications (Cambridge, England)} \textbf{2010}, \emph{46}, 5235--7\relax
\mciteBstWouldAddEndPuncttrue
\mciteSetBstMidEndSepPunct{\mcitedefaultmidpunct}
{\mcitedefaultendpunct}{\mcitedefaultseppunct}\relax
\EndOfBibitem
\bibitem[WINKLER and SINGER(1972)WINKLER, and SINGER]{stonecrystallization}
WINKLER,~E.~M.; SINGER,~P.~C. Crystallization pressure of salts in stone and concrete. \emph{GSA Bulletin} \textbf{1972}, \emph{83}, 3509--3514\relax
\mciteBstWouldAddEndPuncttrue
\mciteSetBstMidEndSepPunct{\mcitedefaultmidpunct}
{\mcitedefaultendpunct}{\mcitedefaultseppunct}\relax
\EndOfBibitem
\bibitem[Shindo \latin{et~al.}(2005)Shindo, Kondo, Shitagami, Sugai, Namai, and Kwak]{Saltfriction2}
Shindo,~H.; Kondo,~S.-i.; Shitagami,~K.; Sugai,~T.; Namai,~Y.; Kwak,~M. Frictional force microscopic detection of anisotropy and asymmetry at various atom-flat surfaces. 2005\relax
\mciteBstWouldAddEndPuncttrue
\mciteSetBstMidEndSepPunct{\mcitedefaultmidpunct}
{\mcitedefaultendpunct}{\mcitedefaultseppunct}\relax
\EndOfBibitem
\bibitem[Socoliuc \latin{et~al.}(2004)Socoliuc, Bennewitz, Gnecco, and Meyer]{saltstickslip}
Socoliuc,~A.; Bennewitz,~R.; Gnecco,~E.; Meyer,~E. Transition from stick-slip to continuous sliding in atomic friction: Entering a new regime of ultralow friction. \emph{Physical Review Letters} \textbf{2004}, \emph{92}, 134301\relax
\mciteBstWouldAddEndPuncttrue
\mciteSetBstMidEndSepPunct{\mcitedefaultmidpunct}
{\mcitedefaultendpunct}{\mcitedefaultseppunct}\relax
\EndOfBibitem
\bibitem[Kim \latin{et~al.}(2010)Kim, Ree, Han, and Shimamoto]{earthquake}
Kim,~J.-W.; Ree,~J.-H.; Han,~R.; Shimamoto,~T. Experimental evidence for the simultaneous formation of pseudotachylyte and mylonite in the brittle regime. \emph{Geology} \textbf{2010}, \emph{38}, 1143--1146\relax
\mciteBstWouldAddEndPuncttrue
\mciteSetBstMidEndSepPunct{\mcitedefaultmidpunct}
{\mcitedefaultendpunct}{\mcitedefaultseppunct}\relax
\EndOfBibitem
\bibitem[Wang \latin{et~al.}(2009)Wang, Mu, Ren, Jian, Zhang, and Yang]{whiskersInPlastic}
Wang,~H.-G.; Mu,~B.; Ren,~J.-F.; Jian,~L.-Q.; Zhang,~J.-Y.; Yang,~S.-R. Mechanical and tribological behaviors of PA66/PVDF blends filled with calcium sulphate whiskers. \emph{Polymer Composites} \textbf{2009}, \emph{30}, 1326--1332\relax
\mciteBstWouldAddEndPuncttrue
\mciteSetBstMidEndSepPunct{\mcitedefaultmidpunct}
{\mcitedefaultendpunct}{\mcitedefaultseppunct}\relax
\EndOfBibitem
\bibitem[{Sudhan Raj} \latin{et~al.}(2020){Sudhan Raj}, Christy, {Darius Gnanaraj}, and Sugozu]{carbrakes}
{Sudhan Raj},~J.; Christy,~T.; {Darius Gnanaraj},~S.; Sugozu,~B. Influence of calcium sulfate whiskers on the tribological characteristics of automotive brake friction materials. \emph{Engineering Science and Technology, an International Journal} \textbf{2020}, \emph{23}, 445--451\relax
\mciteBstWouldAddEndPuncttrue
\mciteSetBstMidEndSepPunct{\mcitedefaultmidpunct}
{\mcitedefaultendpunct}{\mcitedefaultseppunct}\relax
\EndOfBibitem
\bibitem[John \latin{et~al.}(1998)John, Prasad, Voevodin, and Zabinski]{aircraft}
John,~P.; Prasad,~S.; Voevodin,~A.; Zabinski,~J. Calcium sulfate as a high temperature solid lubricant. \emph{Wear} \textbf{1998}, \emph{219}, 155--161\relax
\mciteBstWouldAddEndPuncttrue
\mciteSetBstMidEndSepPunct{\mcitedefaultmidpunct}
{\mcitedefaultendpunct}{\mcitedefaultseppunct}\relax
\EndOfBibitem
\bibitem[John and Zabinski(1999)John, and Zabinski]{John1999}
John,~P.~J.; Zabinski,~J.~S. Sulfate based coatings for use as high temperature lubricants. \emph{Tribology Letters} \textbf{1999}, \emph{7}, 31--37\relax
\mciteBstWouldAddEndPuncttrue
\mciteSetBstMidEndSepPunct{\mcitedefaultmidpunct}
{\mcitedefaultendpunct}{\mcitedefaultseppunct}\relax
\EndOfBibitem
\bibitem[Yamamoto \latin{et~al.}(1994)Yamamoto, Olsson, and Hogmark]{YAMAMOTO199421}
Yamamoto,~T.; Olsson,~M.; Hogmark,~S. Three-body abrasive wear of ceramic materials. \emph{Wear} \textbf{1994}, \emph{174}, 21--31\relax
\mciteBstWouldAddEndPuncttrue
\mciteSetBstMidEndSepPunct{\mcitedefaultmidpunct}
{\mcitedefaultendpunct}{\mcitedefaultseppunct}\relax
\EndOfBibitem
\bibitem[Wang and Hsu(1996)Wang, and Hsu]{WANG1996112}
Wang,~Y.; Hsu,~S.~M. Wear and wear transition mechanisms of ceramics. \emph{Wear} \textbf{1996}, \emph{195}, 112--122\relax
\mciteBstWouldAddEndPuncttrue
\mciteSetBstMidEndSepPunct{\mcitedefaultmidpunct}
{\mcitedefaultendpunct}{\mcitedefaultseppunct}\relax
\EndOfBibitem
\bibitem[Broz \latin{et~al.}(2006)Broz, Cook, and Whitney]{gypsum1}
Broz,~M.~E.; Cook,~R.~F.; Whitney,~D.~L. Microhardness, toughness, and modulus of Mohs scale minerals. \emph{American Mineralogist} \textbf{2006}, \emph{91}, 135--142\relax
\mciteBstWouldAddEndPuncttrue
\mciteSetBstMidEndSepPunct{\mcitedefaultmidpunct}
{\mcitedefaultendpunct}{\mcitedefaultseppunct}\relax
\EndOfBibitem
\bibitem[Zhou \latin{et~al.}(2015)Zhou, Liu, Shu, Yu, Zhang, Li, and Xue]{gypsum2}
Zhou,~J.; Liu,~C.; Shu,~Z.; Yu,~D.; Zhang,~Q.; Li,~T.; Xue,~Q. Preparation of specific gypsum with advanced hardness and bending strength by a novel In-situ Loading-Hydration Process. \emph{Cement and Concrete Research} \textbf{2015}, \emph{67}, 179--183\relax
\mciteBstWouldAddEndPuncttrue
\mciteSetBstMidEndSepPunct{\mcitedefaultmidpunct}
{\mcitedefaultendpunct}{\mcitedefaultseppunct}\relax
\EndOfBibitem
\bibitem[Shahidzadeh-Bonn \latin{et~al.}(2015)Shahidzadeh-Bonn, Schut, Desarnaud, Prat, and Bonn]{dropletevaporation}
Shahidzadeh-Bonn,~N.; Schut,~M.~F.; Desarnaud,~J.; Prat,~M.; Bonn,~D. {Salt stains from evaporating droplets}. \emph{{Scientific Reports}} \textbf{2015}, \emph{vol. 5}, pp. 1--9\relax
\mciteBstWouldAddEndPuncttrue
\mciteSetBstMidEndSepPunct{\mcitedefaultmidpunct}
{\mcitedefaultendpunct}{\mcitedefaultseppunct}\relax
\EndOfBibitem
\bibitem[Popescu \latin{et~al.}(2012)Popescu, Oshanin, Dietrich, and Cazabat]{percussorfilm}
Popescu,~M.~N.; Oshanin,~G.; Dietrich,~S.; Cazabat,~A.-M. Precursor films in wetting phenomena. \emph{Journal of Physics: Condensed Matter} \textbf{2012}, \emph{24}, 243102\relax
\mciteBstWouldAddEndPuncttrue
\mciteSetBstMidEndSepPunct{\mcitedefaultmidpunct}
{\mcitedefaultendpunct}{\mcitedefaultseppunct}\relax
\EndOfBibitem
\bibitem[Qazi \latin{et~al.}(2019)Qazi, Salim, Doorman, Jambon-Puillet, and Shahidzadeh]{creepingtube}
Qazi,~M.~J.; Salim,~H.; Doorman,~C. A.~W.; Jambon-Puillet,~E.; Shahidzadeh,~N. Salt creeping as a self-amplifying crystallization process. \emph{Science Advances} \textbf{2019}, \emph{5}, eaax1853\relax
\mciteBstWouldAddEndPuncttrue
\mciteSetBstMidEndSepPunct{\mcitedefaultmidpunct}
{\mcitedefaultendpunct}{\mcitedefaultseppunct}\relax
\EndOfBibitem
\bibitem[Reitsma \latin{et~al.}(2000)Reitsma, Cain, Smith, Biggs, and Page]{lfm1}
Reitsma,~M.; Cain,~R.; Smith,~D.; Biggs,~S.; Page,~N. Lateral Force Microscopy: A tool for tribology. \emph{GeoEng 2000 Conference} \textbf{2000}, \relax
\mciteBstWouldAddEndPunctfalse
\mciteSetBstMidEndSepPunct{\mcitedefaultmidpunct}
{}{\mcitedefaultseppunct}\relax
\EndOfBibitem
\bibitem[Carpick and Salmeron(1997)Carpick, and Salmeron]{lfm2}
Carpick,~R.~W.; Salmeron,~M. Scratching the Surface: Fundamental investigations of tribology with Atomic Force Microscopy. \emph{Chemical Reviews} \textbf{1997}, \emph{97}, 1163--1194, PMID: 11851446\relax
\mciteBstWouldAddEndPuncttrue
\mciteSetBstMidEndSepPunct{\mcitedefaultmidpunct}
{\mcitedefaultendpunct}{\mcitedefaultseppunct}\relax
\EndOfBibitem
\bibitem[Flater \latin{et~al.}(2018)Flater, Barnes, Hitz~Graff, Weaver, Ansari, Poda, Robert~Ashurst, Khanal, and Jacobs]{quantifytipwear}
Flater,~E.~E.; Barnes,~J.~D.; Hitz~Graff,~J.~A.; Weaver,~J.~M.; Ansari,~N.; Poda,~A.~R.; Robert~Ashurst,~W.; Khanal,~S.~R.; Jacobs,~T. D.~B. A simple atomic force microscope-based method for quantifying wear of sliding probes. \emph{Review of Scientific Instruments} \textbf{2018}, \emph{89}, 113708\relax
\mciteBstWouldAddEndPuncttrue
\mciteSetBstMidEndSepPunct{\mcitedefaultmidpunct}
{\mcitedefaultendpunct}{\mcitedefaultseppunct}\relax
\EndOfBibitem
\bibitem[Wang \latin{et~al.}(2013)Wang, Christenson, and Meldrum]{caso4crystalization1}
Wang,~Y.-W.; Christenson,~H.~K.; Meldrum,~F.~C. Confinement leads to control over calcium sulfate polymorph. \emph{Advanced Functional Materials} \textbf{2013}, \emph{23}, 5615--5623\relax
\mciteBstWouldAddEndPuncttrue
\mciteSetBstMidEndSepPunct{\mcitedefaultmidpunct}
{\mcitedefaultendpunct}{\mcitedefaultseppunct}\relax
\EndOfBibitem
\bibitem[Wang \latin{et~al.}(2012)Wang, Kim, Christenson, and Meldrum]{caso4crystalization2}
Wang,~Y.-W.; Kim,~Y.-Y.; Christenson,~H.~K.; Meldrum,~F.~C. A new precipitation pathway for calcium sulfate dihydrate (gypsum) via amorphous and hemihydrate intermediates. \emph{Chemical Communications} \textbf{2012}, \emph{48}, 504--506\relax
\mciteBstWouldAddEndPuncttrue
\mciteSetBstMidEndSepPunct{\mcitedefaultmidpunct}
{\mcitedefaultendpunct}{\mcitedefaultseppunct}\relax
\EndOfBibitem
\bibitem[Driessche \latin{et~al.}(2012)Driessche, Benning, Rodriguez-Blanco, Ossorio, Bots, and García-Ruiz]{caso4crystalization3}
Driessche,~A. E. S.~V.; Benning,~L.~G.; Rodriguez-Blanco,~J.~D.; Ossorio,~M.; Bots,~P.; García-Ruiz,~J.~M. The role and implications of bassanite as a stable precursor phase to gypsum precipitation. \emph{Science} \textbf{2012}, \emph{336}, 69--72\relax
\mciteBstWouldAddEndPuncttrue
\mciteSetBstMidEndSepPunct{\mcitedefaultmidpunct}
{\mcitedefaultendpunct}{\mcitedefaultseppunct}\relax
\EndOfBibitem
\bibitem[Saha \latin{et~al.}(2012)Saha, Lee, Pancera, Bräeu, Kempter, Tripathi, and Bose]{caso4crystalization4}
Saha,~A.; Lee,~J.; Pancera,~S.~M.; Bräeu,~M.~F.; Kempter,~A.; Tripathi,~A.; Bose,~A. New insights into the transformation of calcium sulfate hemihydrate to gypsum using time-resolved cryogenic transmission electron microscopy. \emph{Langmuir} \textbf{2012}, \emph{28}, 11182--11187, PMID: 22747102\relax
\mciteBstWouldAddEndPuncttrue
\mciteSetBstMidEndSepPunct{\mcitedefaultmidpunct}
{\mcitedefaultendpunct}{\mcitedefaultseppunct}\relax
\EndOfBibitem
\bibitem[Jia \latin{et~al.}(2021)Jia, Wu, Fulton, Liang, De~Yoreo, and Guan]{ACSwater}
Jia,~C.; Wu,~L.; Fulton,~J.~L.; Liang,~X.; De~Yoreo,~J.~J.; Guan,~B. Structural Characteristics of Amorphous Calcium Sulfate: Evidence to the Role of Water Molecules. \emph{The Journal of Physical Chemistry C} \textbf{2021}, \emph{125}, 3415--3420\relax
\mciteBstWouldAddEndPuncttrue
\mciteSetBstMidEndSepPunct{\mcitedefaultmidpunct}
{\mcitedefaultendpunct}{\mcitedefaultseppunct}\relax
\EndOfBibitem
\bibitem[{van Elp} \latin{et~al.}(2004){van Elp}, Giesen, and {de Groof}]{chuck}
{van Elp},~J.; Giesen,~P.; {de Groof},~A. Low-thermal expansion electrostatic chuck materials and clamp mechanisms in vacuum and air. \emph{Microelectronic Engineering} \textbf{2004}, \emph{73-74}, 941--947, Micro and Nano Engineering 2003\relax
\mciteBstWouldAddEndPuncttrue
\mciteSetBstMidEndSepPunct{\mcitedefaultmidpunct}
{\mcitedefaultendpunct}{\mcitedefaultseppunct}\relax
\EndOfBibitem
\bibitem[Hsia \latin{et~al.}(2022)Hsia, Hsu, Peng, Elam, Xiao, Franklin, Bonn, and Weber]{Bartfriction}
Hsia,~F.-C.; Hsu,~C.-C.; Peng,~L.; Elam,~F.~M.; Xiao,~C.; Franklin,~S.; Bonn,~D.; Weber,~B. Contribution of capillary adhesion to friction at macroscopic solid--solid interfaces. \emph{Physical Review Applied} \textbf{2022}, \emph{17}, 034034\relax
\mciteBstWouldAddEndPuncttrue
\mciteSetBstMidEndSepPunct{\mcitedefaultmidpunct}
{\mcitedefaultendpunct}{\mcitedefaultseppunct}\relax
\EndOfBibitem
\bibitem[Thiery \latin{et~al.}(2006)Thiery, Kunze, Karimi, Curnier, and Longchamp]{frictionprecision}
Thiery,~S.; Kunze,~M.; Karimi,~A.; Curnier,~A.; Longchamp,~R. Friction modeling of a high-precision positioning system. 2006; p 5 pp.\relax
\mciteBstWouldAddEndPunctfalse
\mciteSetBstMidEndSepPunct{\mcitedefaultmidpunct}
{}{\mcitedefaultseppunct}\relax
\EndOfBibitem
\bibitem[Liu \latin{et~al.}(1996)Liu, Blanpain, and Celis]{calibration1}
Liu,~E.; Blanpain,~B.; Celis,~J. Calibration procedures for frictional measurements with a lateral force microscope. \emph{Wear} \textbf{1996}, \emph{192}, 141--150\relax
\mciteBstWouldAddEndPuncttrue
\mciteSetBstMidEndSepPunct{\mcitedefaultmidpunct}
{\mcitedefaultendpunct}{\mcitedefaultseppunct}\relax
\EndOfBibitem
\bibitem[Cain \latin{et~al.}(2000)Cain, Biggs, and Page]{calibration2}
Cain,~R.~G.; Biggs,~S.; Page,~N.~W. Force calibration in Lateral Force Microscopy. \emph{Journal of Colloid and Interface Science} \textbf{2000}, \emph{227}, 55--65\relax
\mciteBstWouldAddEndPuncttrue
\mciteSetBstMidEndSepPunct{\mcitedefaultmidpunct}
{\mcitedefaultendpunct}{\mcitedefaultseppunct}\relax
\EndOfBibitem
\end{mcitethebibliography}

\end{document}